\documentclass[12pt, preprint]{aastex}

\newcommand{\rearth}{${\rm R_{\earth}} $}
\newcommand{\mearth}{$\rm M_{\earth}$}

\usepackage{amssymb}
\usepackage{amsmath}
\usepackage{epsfig}
\usepackage{graphicx}
\usepackage{natbib}
\usepackage{multirow}
\usepackage{rotating}
\usepackage{longtable}
\usepackage{verbatim}
\usepackage{url}
\usepackage{hyperref}
\usepackage{caption2}

\begin{document}

\title{
Mass-Radius Relation for Rocky Planets based on PREM
}
\author{Li Zeng\altaffilmark{1,a}, Dimitar Sasselov\altaffilmark{2,b}, and Stein Jacobsen\altaffilmark{1,c} 
}
\affil{$^1$Department of Earth and Planetary Sciences, Harvard University, Cambridge, MA 02138}
\affil{$^2$Astronomy Department, Harvard University, Cambridge, MA 02138}
\email{$^a$astrozeng@gmail.com} 
\email{$^b$dsasselov@cfa.harvard.edu}
\email{$^c$jacobsen@neodymium.harvard.edu} 

\begin{abstract}

Several small dense exoplanets are now known, inviting comparisons to Earth and Venus. Such comparisons require translating their masses and sizes to composition models of evolved multi-layer-interior planets. Such theoretical models rely on our understanding of the Earth's interior, as well as independently derived equations of state (EOS), but have so far not involved direct extrapolations from Earth's seismic model -PREM. In order to facilitate more detailed compositional comparisons between small exoplanets and the Earth, we derive here a semi-empirical mass-radius relation for two-layer rocky planets based on PREM: ${\frac{R}{R_\oplus}} = (1.07-0.21\cdot \text{CMF})\cdot (\frac{M}{M_\oplus})^{1/3.7}$, where CMF stands for Core Mass Fraction. It is applicable to 1$\sim$8 M$_{\oplus}$ and CMF of 0.0$\sim$0.4. Applying this formula to Earth and Venus and several known small exoplanets with radii and masses measured to better than $\sim$30\% precision gives a CMF fit of $0.26\pm0.07$.

\end{abstract}

\section{Introduction}

A first step in deriving the compositional diversity of small rocky planets was accomplished recently by~\citep{Dressing:2015}, who added Kepler-93b to the list of a dozen or so small exoplanets with radii and masses measured to better than $\sim$30\% precision. With exquisite sizes (mostly from Kepler light curves) and huge follow-up effort~\citep{Dressing:2015, Kepler-78b:2013, Batalha:2011, Ballard:2013, Carter:2012, Hatzes:2011, Berta-Thompson:2015} the mass-radius diagram is finally amenable to some detailed comparisons with theory in the 1 to 10~\mearth~range. Closer to the mass and size of Earth, the rocky planets known to-date seem to exhibit an unexpectedly tight compositional correlation. Is this correlation really shared with Earth and Venus? If so, to what level of precision and under what assumptions? To begin answering such questions we must first acknowledge that models of the interior structure and composition of rocky exoplanets in the Super-Earth domain are based largely on experience (and extrapolations) from the models of the rocky solar system planets, and mostly - the Earth~\citep{Valencia:2006, Valencia:2007a, Valencia:2007b, Fortney:2007, Seager:2007, Zeng_Seager:2008, Grasset:2009, Zeng_Sasselov:2013}. Here instead of using simple shell models based on EOSs of minerals and metals, we take a different approach and derive semi-empirical EOSs based on the well-constrained seismic model of the Earth. 

\section{Equation of State (EOS)}

On one hand, in several previous models of solid exoplanets~\citep{Zeng_Sasselov:2013, Zeng_Seager:2008, Seager:2007}, cores and mantles of solid exoplanets are modeled as pure $\epsilon$-Fe-solid and Mg-perovskite/post-perovskite respectively. On the other hand, we know the actual density variation inside Earth through measurements of seismic wave velocities. This seismically derived density model is widely known as PREM (Preliminary Reference Earth Model~\citep{Dziewonski:1981}). The differences between the two approaches, primarily (1) the liquid outer core less dense than pure $\epsilon$-Fe solid and (2) the upper mantle less dense than the extrapolation of lower mantle, can cause differences in the mass-radius relations derived.  

There are several assumptions made when we extrapolate PREM. We assume the upper mantle to lower mantle phase transition occurs at the same pressure as the Earth's interior. It is a pressure-driven phase transition, so temperature effect is secondary and thus ignored. 

The Thomas-Fermi-Dirac Model modified with Correlation Energy~\citep{Salpeter:1967} (abbreviated as TFD from now on) serves as a lower bound for the density of material considered. Any properly-behaving EOS should asymptotically approach TFD above $\sim$1 TPa, as at such high pressures the electron degeneracy pressure dominates while the detailed chemical and crystal structures of the material become less important.

We will show that the Birch-Murnaghan 2nd order EOS (abbreviated as BM2 from now on)~\citep{Birch:1952} provides a fairly decent description of how material is compressed in Earth's interior for both core (good up to 12 TPa where it asymptotically approaches TFD) and mantle (good up to 3.5 TPa where it asymptotically approaches TFD). Those pressures roughly correspond to the central pressure and core-mantle boundary pressure respectively of a 30 M$_\oplus$ rocky planet with CMF=0.3.

\begin{equation}
P=\frac{3}{2} \cdot K_0 \left[\left(\frac{\rho}{\rho_0}\right)^{\frac{7}{3}}-\left(\frac{\rho}{\rho_0}\right)^{\frac{5}{3}} \right]
\label{eq:BM2}
\end{equation}

The fit to lower mantle, outer core and inner core PREM gives: 
\begin{itemize}
\item Lower Mantle: $\rho_0=3.98$ g/cc, $K_0=206$ GPa, error$\sim1\%$ in $\rho$.
\item Outer Core: $\rho_0=7.05$ g/cc, $K_0=201$ GPa, error$\sim1\%$ in $\rho$.
\item Inner Core: $\rho_0=7.85$ g/cc, $K_0=255$ GPa, error$\sim0.01\%$ in $\rho$.
\end{itemize}

The uncertainties of the fit are similar to the intrinsic uncertainty of PREM in $\rho$ of~$\sim1\%$ (See~\citep{Ricolleau:2009} and~\citep{Ritsema:2011} and Figure 3 of~\citep{Huang:2011}).

\subsection{PREM-extrapolated EOS for Mantle}

Inside Earth, the density jump from upper mantle to lower mantle is 10\% from 4 g/cc to 4.4 g/cc at 23.83 GPa. Earth's upper mantle consists of complex phases of Mg-silicates, including various polymorphs of olivine: ($\alpha$) olivine, ($\beta$) wadsleyite, and ($\gamma$) ringwoodite~\citep{Bina:2003}. The upper mantle to lower mantle transition occurs at 670 km depth in Earth (about 10.5\% Earth radius). For more massive planets like Kepler-93b, this transition occurs at shallower depth ($\sim$ 3\% radius of Kepler-93b) for the same pressure.

\begin{figure}
\centering
\includegraphics[width=5.5in]{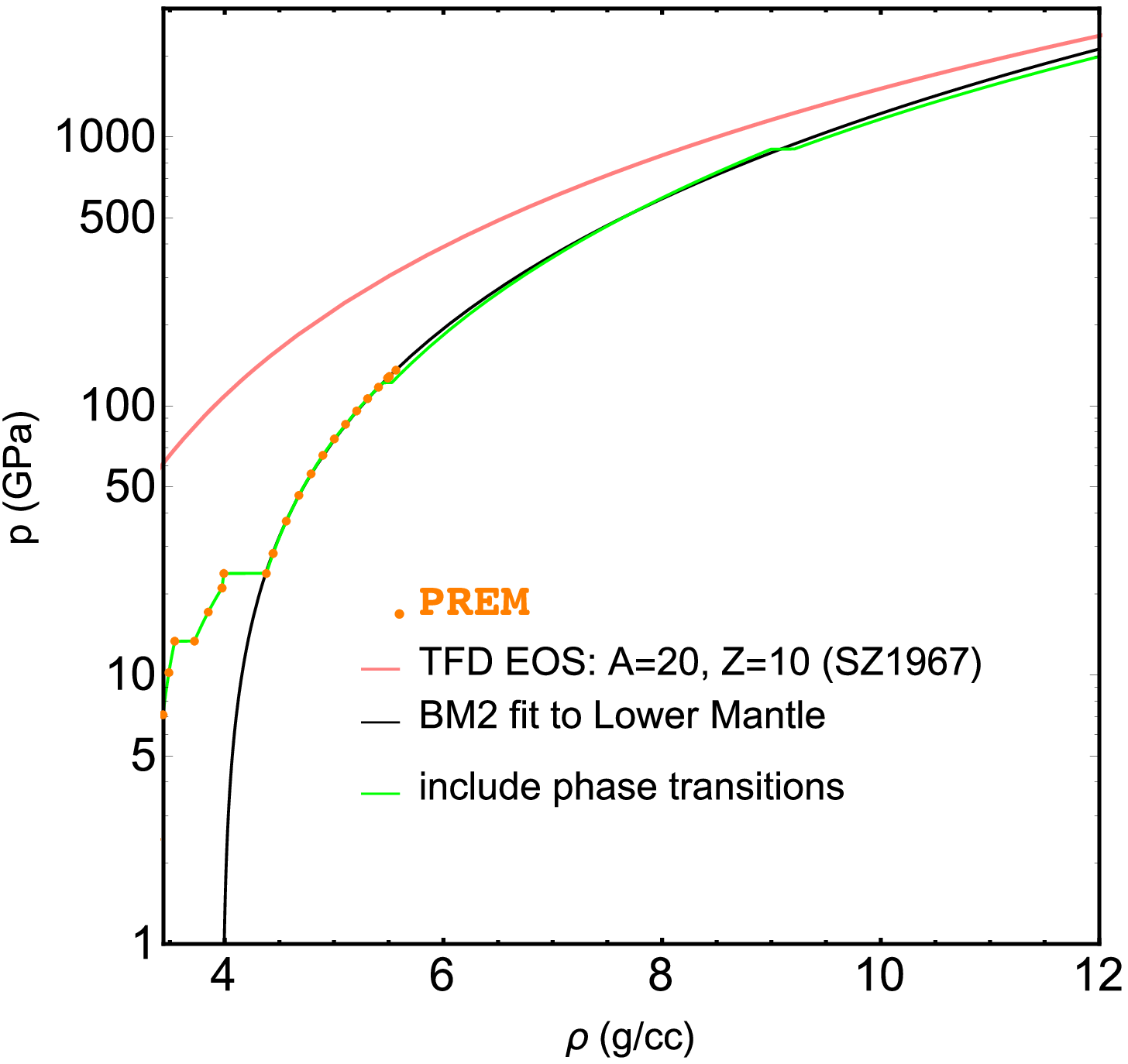}
\caption{\singlespace Orange points: PREM density of Earth's mantle, excerpt from Appendix G of~\citep{Stacey:2008}. Pink curve: TFD for A=20 and Z=10. Black curve: BM2 fit to lower-mantle PREM ($\rho_0=3.98$ g/cc, $K_0=206$ GPa). Green curve: including lower-mantle phase transitions and pv-ppv phase transition at 122 GPa~\citep{Zeng_Sasselov:2013, Caracas:2008, Spera:2006} and further dissociation of ppv at 0.9 and 2.1 TPa~\citep{Umemoto:2011, Wu:2011}.}
\label{eosmantle}
\end{figure}

Fig.~\ref{eosmantle} shows the comparison of different EOS of Mg-silicates.

Interestingly, the TFD beyond 1 TPa for MgO, SiO$_2$, MgSiO$_3$, and Mg$_2$SiO$_4$ are almost identical as they all have average atomic weight A=20 and average atomic charge Z=10. This coincidence simplifies the EOS of Mg-silicates, as it indicates that Mg/Si ratio does not matter towards the high-pressure end. It only matters towards the low-pressure end, which is captured in our EOS by using the PREM density variation in the upper-mantle pressure range.

According to Fig.~\ref{eosmantle}, except the prominent density jump of 10\% at the upper-lower mantle boundary, the other high-pressure phase transitions are only $\sim$ 1\% level in density and thus insignificant in mass-radius calculation~\citep{Unterborn:2015}. Therefore, in this paper we adopt the EOS of Mg-silicates mantle as follows:

\begin{itemize}
\item 0 GPa - 23.83 GPa: linear interpolation of the lower mantle density according to the Appendix G of~\citep{Stacey:2008}, which is taken from PREM. i.e. follow the green curve in Fig.~\ref{eosmantle}.
\item 23.83 GPa - 3.5 TPa: BM2 EOS with $\rho_0=3.98$ g/cc, $K_0=206$ GPa, (error$\sim1\%$ in density). i.e. follow the black curve in Fig.~\ref{eosmantle}.
\item $>$3.5 TPa: TFD EOS of MgSiO$_3$ calculated using method in~\citep{Salpeter:1967}. i.e. follow the pink curve in Fig.~\ref{eosmantle}.
\end{itemize}

\subsection{PREM-extrapolated EOS for Core}

\begin{figure}
\centering
\includegraphics[width=5.5in]{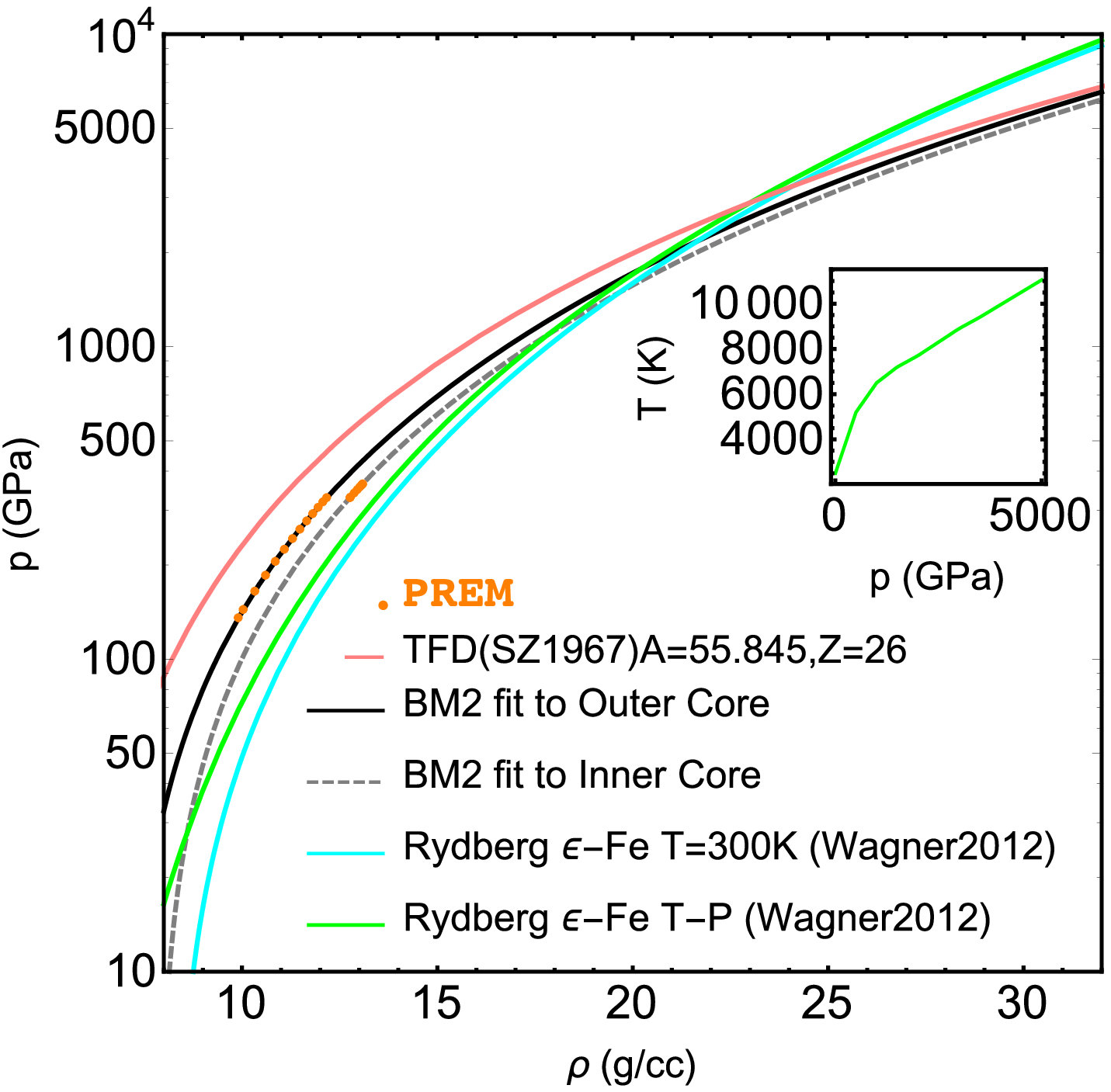}
\caption{\singlespace Orange points: PREM density of Earth's core, excerpt from Appendix G of~\citep{Stacey:2008}. Pink curve: TFD of iron (A=55.845 and Z=26). Black curve: BM2 fit to outer-core PREM ($\rho_0=7.05$ g/cc, $K_0=201$ GPa). Gray dashed curve: BM2 fit to inner-core PREM ($\rho_0=7.85$ g/cc, $K_0=255$ GPa). Cyan curve: Rydberg $\epsilon$-Fe EOS at 300K isotherm according to~\citep{Dewaele:2006}. Green curve: Rydberg $\epsilon$-Fe EOS~\citep{Dewaele:2006} with T-P profile (small inlet plot) of rocky planet core interpolated from~\citet{Wagner:2012}.}
\label{eoscore}
\end{figure}

Fig.~\ref{eoscore} shows the comparison of different EOS of the Core.

In Earth, the density jump from liquid outer core to solid inner core is 5\% from 12.2 g/cc to 12.8 g/cc at 328.85 GPa. The Rydberg EOS of pure solid $\epsilon$-Fe phase, which is experimentally determined from static compression up to 205 GPa~\citep{Dewaele:2006, Wagner:2012}, is too high of density (cyan curve in Fig.~\ref{eoscore}) in the pressure region relevant to Earth, even including the temperature effect (green curve in Fig.~\ref{eoscore}), even for the solid inner core. This is likely due to the presence of several weight percent of one or more light elements~\citep{McDonough:2014}. The lighter elements could be S, Si, O, C, or a combination of them~\citep{Badro:2007, Fischer:2012, Alfe:2002, Poirier:1994, Stixrude:1997, Anderson:1994}, but no agreement is reached as to which of these elements are most important. The Rydberg EOS with or without temperature behaves poorly for extrapolation above$\sim$1TPa as it overshoots and crosses over the TFD EOS, which serves as a lower bound for the density of Fe.

As such, we choose to only extrapolate the liquid outer core BM2 EOS to higher pressure until it asymptotically approaches the TFD EOS. The solid inner core is likely a secondary feature resulting from the crystallization of the liquid core. It currently comprises only a small fraction ($\approx 3 \%$) of the total mass of the Earth, and in more massive planets, the focus of this study, solid inner core may not exist at all due to higher heat content and slower cooling rate. BM2 EOS behaves nicely as it asymptotically approaches the TFD EOS at vert high pressure ($\sim$12 TPa).

Therefore, in this paper we adopt the EOS of Fe core as follows:

\begin{itemize}
\item 0 GPa - 12 TPa: BM2 EOS with $\rho_0=7.05$ g/cc, $K_0=201$ GPa, (error$\sim1\%$ in density). i.e. follow the black curve in Fig.~\ref{eoscore}.
\item $>$12 TPa: TFD EOS of Fe (A=55.845 and Z=26) calculated using method in~\citep{Salpeter:1967}. i.e. follow the pink curve in Fig.~\ref{eoscore}.
\end{itemize}

\section{Mass-Radius Relation}

\begin{figure}[!htb]
\centering
\includegraphics[width=5.0in]{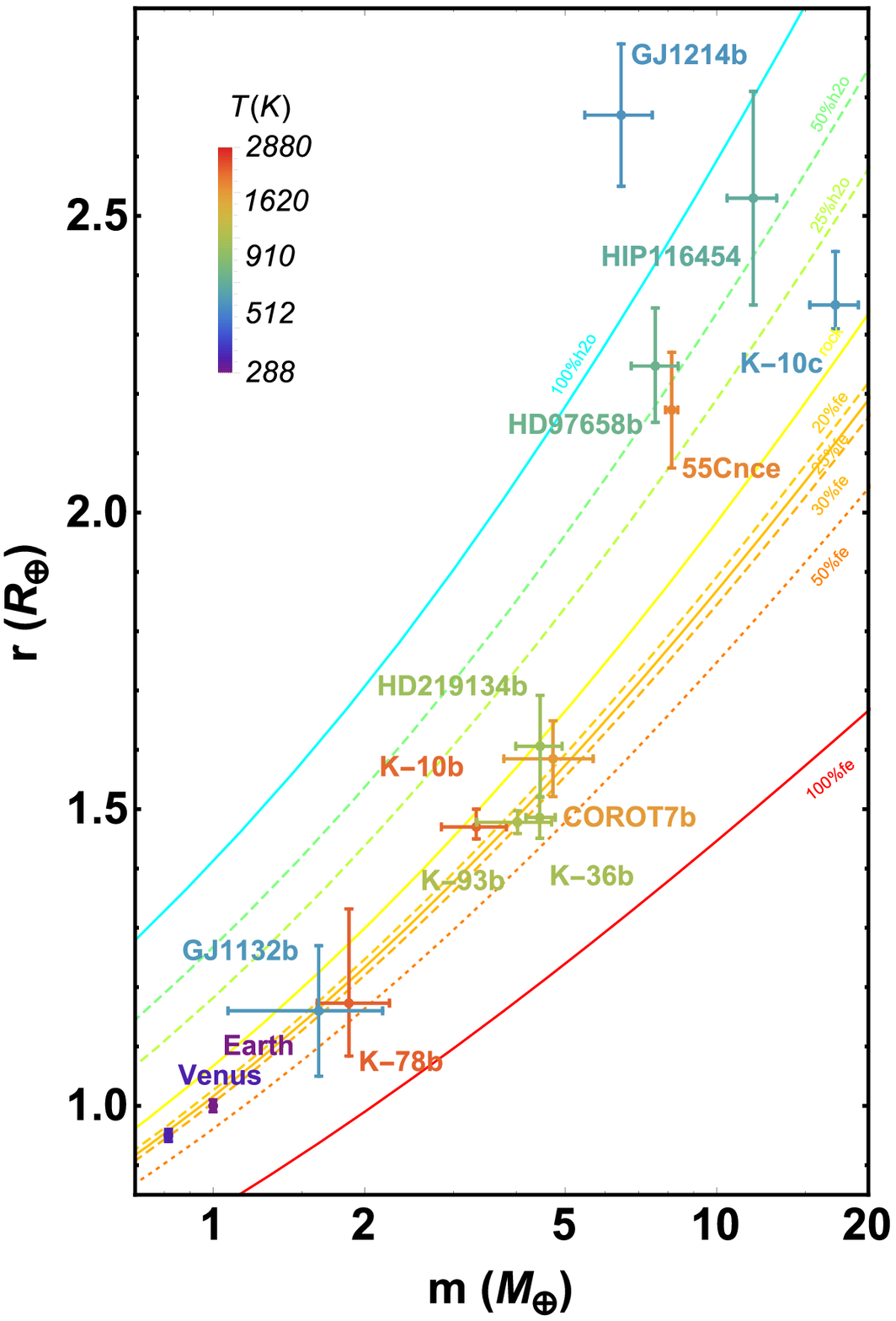}
\caption{\singlespace Mass-radius curves with planets color-coded by their surface temperatures (calculated from the stellar flux they receive assuming similar bond albedo as the Earth ($\approx$0.3) and perfect heat redistribution).~\citep{Kepler-78b:2013, Dumusque:2014, Dressing:2015, Ballard:2013, Carter:2012, Haywood:2014, Barros:2014, Dragomir:2013, Nelson:2014, Winn:2011, Vanderburg:2015, Cochran:2011, Charbonneau:2009, Motalebi:2015, Berta-Thompson:2015}}
\label{mrplot}
\end{figure}

\cite{Dressing:2015} points out a tight mass-radius relation of solid exoplanets between 2 and 5~\mearth~from the comparison of Earth, Venus, and several dense exoplanets by using the mass-radius contours described in~\citep{Zeng_Sasselov:2013}. However, \cite{Dressing:2015}'s CMF fit of 17\% is much lower than that of Earth or Venus. Here we show that using the PREM-extrapolated EOS renders a better fit of CMF in agreement with that of Earth and Venus (See Fig.~\ref{mrplot}).  

There might be a selection bias towards higher density planets by selecting only the planets with better than $\sim$30\% mass measurement accuracy. However, as already pointed out by~\citet{Dressing:2015}, the low mass planets with very low densities like Kepler-11b and Kepler-138d do not detract from the conclusion that all the rocky analogs of the Earth obey a single mass-radius relation. The degree to which this is true will be tested by more precise mass measurements. 

The planets in Fig.~\ref{mrplot} are tabulated in Table~\ref{Table1}.

\begin{table}[htbp]

		\caption{CMF of Planets}

			\centering
			\scalebox{0.8}{
			
			\begin{tabular}{| c | c | c | c |}
			
			\hline

				Planet & M(M$_{\oplus}$) & R(R$_{\oplus}$) & CMF \\

				\hline

				Earth\footnotemark[1] & $1$ & $1$ & $0.325\pm0.001$ \\
				\hline
				Venus\footnotemark[1] & $0.815$ & $0.9499$ & $0.31\pm0.01$ \\
				\hline
				K-10b & $3.33\pm0.49$~\citep{Dumusque:2014} & $1.47_{-0.02}^{+0.03}$~\citep{Dumusque:2014} & $0.04\pm0.2$ \\
				\hline
				K-36b & $4.45_{-0.27}^{+0.33}$~\citep{Carter:2012} & $1.486\pm0.035$~\citep{Carter:2012} & $0.37\pm0.14$  \\
				\hline
				K-78b & $1.86_{-0.25}^{+0.38}$~\citep{Kepler-78b:2013} & $1.173_{-0.089}^{+0.159}$~\citep{Kepler-78b:2013} & $0.37^{+0.3}_{-0.5}$  \\
				\hline
				K-93b & $4.02\pm0.68$~\citep{Dressing:2015} & $1.478\pm0.019$~\citep{Ballard:2013} & $0.26\pm0.2$  \\
				\hline
				COROT-7b & $5.74\pm0.86$~\citep{Haywood:2014} & $1.585\pm0.064$~\citep{Barros:2014} & $0.14\pm0.3$ \\
				\hline
				HD219134b & $4.46\pm0.47$~\citep{Motalebi:2015} & $1.606\pm0.086$~\citep{Motalebi:2015} & $0.0\pm0.3$ \\
				\hline 
				GJ1132b & $1.62\pm0.55$~\citep{Berta-Thompson:2015} & $1.16\pm0.11$~\citep{Berta-Thompson:2015} & $0.25\pm0.5$ \\
				\hline 
			
		  \end{tabular}
	}	
	 \label{Table1}	
\end{table}
\footnotetext[1]{In the calculation, we treat the errors in CMF of Earth and Venus as $\pm0.2$, comparable to the errors for exoplanets considered here, so as not to bias the fit.}

The mass-radius curves in Fig.~\ref{mrplot} are tabulated in Table~\ref{Table2}.

\begin{table}[htbp]
\caption{Mass-Radius Table}

\centering
\scalebox{1.0}{
\begin{tabular}{c c c c c c c c c c}
\\
\hline\hline

 & 100\%fe & 50\%fe & 30\%fe & 25\%fe & 20\%fe & rock & 25\%h2o & 50\%h2o & 100\%h2o \\ \hline
 
M(\mearth)\footnotemark[2] & R(\rearth)\footnotemark[3] & R(\rearth) & R(\rearth) & R(\rearth) & R(\rearth) & R(\rearth) & R(\rearth) & R(\rearth) & R(\rearth) \\ \hline

0.125 &	0.445 &	0.523 &	0.547 &	0.553 &	0.558 &	0.58  &	0.649 &	0.697 &	0.776  \\
0.25  &	0.55  &	0.645 &	0.672 &	0.679 &	0.685 &	0.711 &	0.793 &	0.851 &	0.952  \\
0.5 &  	0.676 &	0.789 &	0.823 &	0.832 &	0.84  &	0.872 &	0.969 &	1.039 &	1.163  \\
1   &  	0.823 &	0.961 &	1.005 &	1.016 &	1.026 &	1.067 &	1.182 &	1.27  &	1.41   \\
2  &   	0.99  &	1.164 &	1.22  &	1.23  &	1.25  &	1.3  & 	1.44  &	1.54  &	1.71   \\
4   &  	1.176 &	1.4 &  	1.47  &	1.49  &	1.5  & 	1.57  &	1.74  &	1.85  &	2.05   \\
8   &  	1.38 & 	1.66 & 	1.75  &	1.77  &	1.79  &	1.88  &	2.07  &	2.21  &	2.45   \\
16   & 	1.59  &	1.94  &	2.06  &	2.08 & 	2.11 & 	2.22  &	2.45  &	2.61  &	2.9    \\
32  &  	1.82  &	2.25  &	2.38  &	2.42  &	2.45  &	2.58  &	2.85  &	3.04  &	3.36   \\

\hline
\end{tabular}
}
\label{Table2}
\end{table}

\footnotetext[2]{mass in Earth Mass ($M_{\oplus} = 5.9742\times10^{24} kg$)}
\footnotetext[3]{radius in Earth Radius ($R_{\oplus} = 6.371\times10^{6} m$)}

\subsection{Mass-Radius Formula Fitting for Rocky Planets}

The CMF-dependent mass-radius relation of rocky planets can be fit to the following formula (for 0$\leq$CMF$\leq$0.4, $1M_{\oplus}\leq M \leq8M_{\oplus}$). It agrees with the actual mass-radius curves in Fig.~\ref{mrplot} within $\sim 0.01~R_{\oplus}$ in radius. If we use it to calculate CMF, it gives an uncertainty of $\sim0.02$ in CMF. 

\begin{equation}
\left( {\frac{R}{{{R_ \oplus }}}} \right) = (1.07-0.21\cdot \text{CMF}) {\left( {\frac{M}{{{M_ \oplus }}}} \right)^{1/3.7}}
\label{eq:mr}
\end{equation}

Eq.~\ref{eq:mr} can be inverted to solve for CMF given the mass and radius: 

\begin{equation}
\text{CMF} = \frac{1}{0.21} [1.07-(\frac{R}{R_\oplus})/(\frac{M}{M_\oplus})^{1/3.7}]
\label{eq:cmf}
\end{equation}

For comparison, this semi-empirical mass-radius formula is in agreement with the mass-radius relation of super-Earths presented in~\citet{Valencia:2006}, which is a scaling law of $R \propto M^{0.267-0.272}$. This new formula takes one step further to articulate the dependence on CMF, which is useful in differentiating among rocky planets with different CMF (i.e. compositions).

For Earth,  CMF$_{\oplus} = \frac{1}{0.21} [1.07-(1)/(1)^{1/3.7}] = 0.07/0.21 = 0.33$. 
For Kepler-93b, CMF$_{K93b} = \frac{1}{0.21} [1.07-(1.478)/(4.02)^{1/3.7}] = 0.26$. 
The uncertainties ($\delta$CMF) in CMF resulting from the uncertainties in mass and radius can be derived from Eq.~\ref{eq:cmf}: 

\begin{equation}
|\delta\text{CMF}| \approx 5\times \sqrt{|\frac{\delta r}{r}|^2+(\frac{1}{3.7})^2|\frac{\delta m}{m}|^2}
\label{eq:cmferror}
\end{equation}

So in order to tell a 20\% core mass apart from a 30\% core mass, radius needs to be measured to better than 2\% level, or equivalently, mass to 6\% level. 
For Kepler-93b, $|\frac{\delta r}{r}| = \frac{0.019}{1.478} = 0.013$ and $|\frac{\delta m}{m}| = \frac{0.68}{4.02} = 0.17$, so $|\delta\text{CMF}| = 0.2$. Therefore, CMF$_{K93b} = 0.26\pm0.2$. Table~\ref{Table1} Column 4 lists the CMF estimates of these exoplanets.  

Assuming this population of dense exoplanets follows the same CMF distrbution, we then apply a weighted least-square fit to Table~\ref{Table1}.

Denote the weighted average of CMF as $\overline{\text{CMF}}$: 

\begin{equation}
\overline{\text{CMF}} = \frac{\sum\limits_{i} \left( \frac{\text{CMF}}{{\delta \text{CMF}}^2} \right)}{\sum\limits_{i} \left( \frac{1}{{{\delta \text{CMF}}^2}} \right)}
\label{eq:weighted-average}
\end{equation}

Denote the standard deviation of CMF as $\sigma \text{CMF}$: 

\begin{equation}
\sigma \text{CMF} = \sqrt {1/ \left( \sum\limits_{i}\frac{1}{{{\delta \text{CMF}}^2}} \right)}
\label{eq:weighted-average-error}
\end{equation}

Result: $\overline{\text{CMF}}=0.26$ and $\sigma \text{CMF}=0.07$, indicating Earth-like composition ($\text{CMF}\sim0.3$) carries on up to at least$~\sim$5 M$_{\oplus}$.

\subsection{Discussion}

This is backed up by recent studies of disintegrated planet debris in polluted white dwarf spectra~\citep{Jura:2014, Wilson:2015, Xu:2014}. These studies show that the accreted extrasolar planet debris generally resemble bulk Earth composition ($>85\%$ by mass composed of O, Mg, Si, Fe), similar Fe/Si and Mg/Si ratio and are carbon-poor. It indicates formation processes similar to those controlling the formation and evolution of objects in the inner solar system~\citep{Jura:2014}.

In our solar system, evidence suggests that rocky bodies were formed from chondritic-like materials~\citep[cf.][]{Lodders:2010}. Current planet formation theory suggests that the solar nebula was initially heated to very high temperatures to the extent that virtually everything was vaporized except for small amount of presolar grains~\citep{Ott:2007}. The nebula then cools to condense out various elements and mineral assemblages from the vapor phase at different temperatures according to the condensation sequence~\citep{White:2013}. Fe-Ni metal alloy and Mg-silicates condense out around similar temperatures of 1200-1400K (depending on the pressure of the nebula gas) according to thermodynamic condensation calculation~\citep{Lodders:2003}. Oxygen, on the other hand, does not have a narrow condensation temperature range, as it is very abundant and it readily combines with all kinds of metals to form oxides which condense out at various temperatures as well as H, N, C to form ices condensing out at relatively low temperatures~\citep{Lewis:1997}. As supported by the polluted white dwarf study, we expect other exoplanetary systems to follow similar condensation sequence as the solar system in a H-dominated nebular environment for the major elements: Fe, Mg, Si, and O~\citep{Jura:2014}.

In solar system, the primitive CI Carbonaceous Chondrites have $\text{Fe}:\text{Mg}:\text{Si} = 0.855 : 1.047 : 1$~\citep{McDonough:1995}. If all this Fe forms the core, CMF$\approx0.38$ is the upper limit. If some Fe is incorporated into the mantle either as metal or oxides, CMF becomes less. The solar ratio of Fe/Si is representative for the stars in the solar neighborhood, which is a tight distribution centered at 1, while the Mg/Si=1 seems to tend towards the lower end of the distribution centered at 1.34~\citep{Gilli:2006, Grasset:2009}. A Mg/Si ratio higher than 1 could produce more olivine (Mg$_2$SiO$_4$) or more MgO to affect the mineralogy of the upper mantle~\citep{Bond:2010, Delgado-Mena:2010, Pagano:2015}. However, it does not affect high-pressure EOS much.  As we have pointed out earlier in Section 2 of EOS, the TFD EOS of MgO, SiO$_2$, MgSiO$_3$, Mg$_2$SiO$_4$ will converge above $\sim$1TPa due to their identical average atomic weight of 20 and atomic charge of 10. Therefore, for more massive planets the effect of Mg/Si tends to be smaller. The dispersion expected from the variation in Mg/Si and Fe/Si ratios cause approximately 2\% difference in radius~\citep{Grasset:2009, Dressing:2015}. 

Oxygen is more readily available than Fe, Mg, or Si~\citep{Lodders:2003}, as it is richly produced in nuclear synthesis of massive stars and chemical evolution of Galaxy~\citep{Pagel:1997}. So there is usually enough O to combine with Mg and Si to form Mg-silicates as well as to oxidize some Fe-metal. Comparing bodies in our solar system: the core mass fractions of Earth and Venus~\citep{Rubie:2007} are around 0.3, the core mass fraction of Mars is estimated to be 0.2~\citep{McSween:2003}, and of Vesta is smaller, about 0.17~\citep{Ermakov:2014, Ruzicka:1997}. For asteroid parent bodies of iron meteorites, their core mass fractions are even smaller~\citep{Petaev:2004}. As such, there seems to exist a trend of increasing CMF from smaller objects towards bigger objects in the solar system. This trend seems to turn flat around 1~\mearth. 

These dense exoplanets between 2 and 5~\mearth~so far appear to agree with the mass-radius relation with $\text{CMF}\approx0.26$, suggesting that they are like the Earth in terms of their proportions of mantle and core. But their surface conditions are utterly different as they are much too hot. This is due to observational bias that currently it is much easier for us to detect close-in planets around stars. The fact that we now see so many of them suggests there may be abundant Earth-like analogs at proper distances from their stars to allow existence of liquid water on their surfaces.

\section{Summary}

This paper provides a new CMF-dependent semi-empirical mass-radius relation for rocky planets of 0$\leq$CMF$\leq$0.4 and $1M_{\oplus}\leq M \leq8M_{\oplus}$.

The result of fit to several dense exoplanets: CMF$\approx0.26\pm0.07$, agrees with the studies of disintegrated planet debris in polluted white dwarf spectra, the solar system formation theory and geochemical and cosmochemical evidence of meteorites.

The model tool is accessible at \url{www.astrozeng.com}.

\section{Acknowledgement}
The author Li Zeng is grateful to Simons Foundation for supporting his postdoctoral research position under the Simons Collaboration on the Origins of Life. 
The author Li Zeng would like to thank Courtney Dressing, Lars Buchhave, and Eugenia Hyung for inspiring discussions.  
Part of this research was conducted under the Sandia Z Fundamental Science Program and supported by the Department of Energy National Nuclear Security Administration under Award Number DE-NA0001804 to S. B. Jacobsen (PI) with Harvard University. This research is the authors' views and not those of the DOE.

\clearpage


\bibliographystyle{apj}
\bibliography{mybib}

\end{document}